\documentclass{emulateapj}
\pdfoutput=1

\usepackage{natbib}
\shorttitle{M82 \hcn/HCO+ with the GBT}
\shortauthors{Kepley et al.}

\newcommand{\kms}{\ensuremath{\rm{km \, s}^{-1}}}

\newcommand{\cc}{\ensuremath{\rm{cm^{-3}} \ }}
\newcommand{\co}[3]{\ensuremath{\rm{^{#1} CO ({#2}-{#3})}}}
\newcommand{\Kkms}{\ensuremath{\rm{K \, km \, s^{-1}}}}

\newcommand{\hcn} {\ensuremath{{\rm HCN}}}
\newcommand{\hcop}{\ensuremath{{\rm HCO^+}}}
\newcommand{\LIR}{\ensuremath{{\rm L_{IR}}}}
\newcommand{\LCO}{\ensuremath{{\rm L_{CO}}}}
\newcommand{\LHCN}{\ensuremath{{\rm L_{HCN}}}}
\newcommand{\LHCOP}{\ensuremath{{\rm L_{HCO+}}}}
\newcommand{\LCOHCN}{\ensuremath{L_{CO}/L_{HCN}}}
\newcommand{\LHCNCO}{\ensuremath{L_{HCN}/L_{CO}}}
\newcommand{\LCOHCOP}{\ensuremath{L_{CO}/L_{HCO+}}}
\newcommand{\LHCOPCO}{\ensuremath{L_{HCO+}/L_{CO}}}
\newcommand{\LHCNHCOP}{\ensuremath{L_{HCN}/L_{HCO+}}}

\begin{document}


\title{The Green Bank Telescope Maps the Dense, Star-Forming Gas in the Nearby
  Starburst Galaxy M82}

\author{Amanda A. Kepley}
\affil{National Radio Astronomy Observatory, P.O. Box 2, Green Bank,
  WV 24944-0002}
\email{akepley@nrao.edu}

\author{Adam K. Leroy}
\affil{National Radio Astronomy Observatory, 520 Edgemont Road,
  Charlottesville, VA 22903-2475}

\author{David Frayer}
\affil{National Radio Astronomy Observatory, P.O. Box 2, Green Bank,
  WV 24944-0002}

\author{Antonio Usero}
\affil{Observatorio Astron{\'o}mico Nacional, C/ Alfonso XII, 3,
  E-28014 Madrid, Spain}

\author{Josh Marvil} \affil{Department of Physics, New Mexico Tech.,
  801 Leroy Place, Socorro, NM 87801, USA ; National Radio Astronomy
  Observatory, P.O. Box O, 1003 Lopezville Road, Socorro, NM 87801,
  USA}

\author{Fabian Walter}
\affil{Max Planck Institute fur Astronomie, K{\"o}nigstuhl 17, D-69117
  Heidelberg, Germany}

\begin{abstract}

  Observations of the Milky Way and nearby galaxies show that dense
  molecular gas correlates with recent star formation, suggesting that
  the formation of this gas phase may help regulate star formation. A
  key test of this idea requires wide-area, high-resolution maps of
  dense molecular gas in galaxies to explore how local physical
  conditions drive dense gas formation, but these observations have
  been limited because of the faintness of dense gas tracers like
  \hcn\ and \hcop. Here we demonstrate the power of the Robert C. Byrd
  Green Bank Telescope -- the largest single-dish millimeter radio
  telescope -- for mapping dense gas in galaxies by presenting the
  most sensitive maps yet of \hcn\ and \hcop\ in the starburst galaxy
  M82.  The \hcn\ and \hcop\ in the disk of this galaxy correlates
  with both recent star formation and more diffuse molecular gas and
  shows kinematics consistent with a rotating torus.  The \hcop\
  emission extending to the north and south of the disk is coincident
  with the outflow previously identified in CO and traces the eastern
  edge of the hot outflowing gas. The central starburst region has a
  higher ratio of star formation to dense gas than the outer regions,
  pointing to the starburst as a key driver of this relationship.
  These results establish that the GBT can efficiently map the dense
  molecular gas at 90~GHz in nearby galaxies, a capability that will
  increase further with the 16 element feed array under construction.

%



\end{abstract}



\section{Introduction} \label{sec:introduction}


Observations of local Milky Way clouds show that the bulk of molecular
gas is inert with star formation concentrated in the small fraction of
the cloud at high surface density
\citep[][]{2010ApJ...723.1019H,2010ApJ...724..687L,2012ApJ...745..190L}. This
observation leads to the idea that the ``dense'' molecular gas ($A_V
\gtrsim 10$~mag, $n(H_2)\gtrsim 10^4 \, \cc$), rather than the whole
molecular interstellar medium ($A_V \gtrsim 1$~mag, $n(H_2)\gtrsim
10^2 \, \cc$), represents the star-forming phase.

Extragalactic scaling relations also support the idea that the amount
of high density molecular gas helps to set the star formation rate.
Low angular resolution spectroscopy of nearby galaxies reveal a
constant ratio (within a factor of a few) of HCN intensity, which
traces the mass of dense gas, to infrared (IR) emission, which traces
the rate of recent star formation, across a wide range of systems,
including the central parts of nearby disks, starbursts, and major
mergers
\citep{2004ApJ...606..271G,2009ApJ...707.1217J,2012A&A...539A...8G}. Surprisingly,
the HCN-to-IR ratios for whole galaxies are comparable to those in
local cloud cores \citep{2010ApJS..188..313W}, implying a constant
ratio of dense gas to star formation rate. The ratios of CO intensity,
which trace the overall molecular gas mass, to the IR emission,
however, vary by more than a factor of ten between disk and starburst
galaxies, suggesting that in extreme regions the relationship between
the total molecular gas mass and the amount of star formation may be
non-linear
\citep{2004ApJ...606..271G,2013AJ....146...19L,2013ARA&A..51..105C}.

If the amount of dense gas is linked to the amount of star formation,
then the formation of dense gas from more diffuse molecular gas
represents an important regulating process. While there may be many
regulating steps (e.g., the formation of giant molecular clouds out of
diffuse {\sc Hi}, accretion onto the galaxy), differences among
density probability distribution functions (PDFs) of local clouds
\citep{2009A&A...508L..35K} and the contrast between ULIRGs, LIRGs,
and normal spirals \citep[][Usero et al., in
prep.]{2012A&A...539A...8G} support the idea that the ratio of dense
to total molecular gas varies in different environments.

This hypothesis can be tested in nearby galaxies by comparing dense
gas, as traced by \hcn\ and \hcop, to the total molecular gas, as
traced by CO.  While observations of nearby galaxies have lower
spatial resolution and sensitivity than observations in the Milky Way,
nearby galaxies probe a wider range of physical conditions than found
in the Solar Neighborhood and more directly relate positions with
environmental conditions. The main obstacle to pursuing systematic
studies of the \hcn-to-CO ratio, or analogous measures of the dense
gas fraction like the \hcop-to-CO ratio, has been the faintness of the
line emission: averaged over a large part of a galaxy, the \hcn\ and
\hcop\ lines are 10--30 times fainter than $\rm{CO}$
\citep{2004ApJ...606..271G}.  As a result, most studies of
extragalactic dense gas with small-aperture millimeter-wave telescopes
have focused on individual deep pointings rather than wide-field
maps. The new $\lambda \sim 4$mm (``W band'') heterodyne receiver on
the Robert C. Byrd Green Bank Telescope (GBT) has the potential to
extend these single-pointing observations and map \hcn\ and \hcop\
distributions in nearby galaxies because of the GBT's large collecting
area, high surface accuracy, and good resolution.

In this Letter, we demonstrate the power of the GBT as an \hcn\ and
\hcop\ mapping machine for nearby galaxies using new GBT maps of
$\hcn(J = 1-0)$ and $\hcop(J=1-0)$ in the nearby
\citep[D=3.530~Mpc;][]{2009ApJS..183...67D} starburst galaxy M82
\citep{1980ApJ...238...24R}.  These observations -- from less than 15
hours of telescope time -- have the best surface brightness
sensitivity of any published \hcn\ or \hcop\ map of M82 and resolution
$< 200$~pc.  We compare the \hcn\ and \hcop\ emission to that of the
bulk molecular gas traced by \co{12}{2}{1} and \co{12}{1}{0} and to
recent star formation traced by 3~GHz radio continuum emission.

\section{GBT Observations}
\label{sec:data}

\begin{deluxetable*}{lccc}
\tablewidth{0pt}
\tabletypesize{\scriptsize}
\tablecaption{Summary of Observations \label{tab:obs_summary}}
\tablehead{
\colhead{} &
\colhead{9 March 2013} &
\colhead{10 March 2013} &
\colhead{21 April 2013}
}
\startdata
UT Times                        &  3:15 -- 11:15   & 3:45 -- 6:45  & 2:00 -- 5:45 \\
Number of Spectral Windows      &       2          &       2       &    2         \\
Bandwidth per Window (MHz)      &    200           &     200       & 200 \\
Spectral Resolution (kHz)       &  24.414          &    12.207     & 12.207 \\
Flux Calibrator & 0359+5057 & 1229+0203 & 1229+0203 \\
Flux Calibrator Flux (Jy)\tablenotemark{a} & 4.0\tablenotemark{b} & 8.5\tablenotemark{b} &  9.9\tablenotemark{c} \\
Pointing Source & 0841+7053 & 0841+7053 & 0841+7053 \\
Average Opacity\tablenotemark{d} & 0.086 & 0.109 & 0.082 \\
Average System Temperature (\hcn, \hcop) (K) & 109, 117 & 112, 117 & 89, 88 \\
Main Beam Efficiency (\hcn, \hcop) & 0.26, 0.29 & 0.26, 0.30 & 0.23,0.22\tablenotemark{e} \\
\enddata
\tablenotetext{a}{Values from the CARMA Calfind database
  (\url{http://carma.astro.umd.edu/cgi-bin/calfind.cgi}). }
\tablenotetext{b}{At 103.5 GHz.}  \tablenotetext{c}{At 92.6 GHz.}
\tablenotetext{d}{From GBT High Frequency Weather Forecasts
  (\url{http://www.gb.nrao.edu/~rmaddale/WeatherNAM/}).}
\tablenotetext{e}{The actuators for a section of panels were
  inoperational for this session leading to lower efficiencies.}
\end{deluxetable*}

We used the newly commissioned 4mm (``W band'') receiver and the GBT
Spectrometer to simultaneously map the $\hcn(J=1-0)$ ($\nu_{rest}$ =
88.63160~GHz) and $\hcop(J=1-0)$ ($\nu_{rest}$ = 89.18853~GHz)
intensity across the central 1.75\arcmin\ by 1.5\arcmin\ ($1.8 \times
1.5$~kpc) of M82. Table~\ref{tab:obs_summary} gives the details of the
observations (GBT project 13A-253).





We used a single beam of the W-band receiver to make rapid maps,
sampling five times each beam FWHM in the scanning direction and
sampling 2.4 times each beam FWHM in the orthongonal direction
\citep{2007A&A...474..679M}.




The observations were made over $\sim15$h in excellent
weather. Out-of-focus holography scans were made every two hours to
correct the dish surface, pointing and focus checks were made every
hour, and flux calibration observations were done every observing
session. The receiver gain was calibrated using hot and cold
loads.\footnote{See http://www.gb.nrao.edu/4mm/.}  The ``OFF''
spectrum was created using the average of the first and last
integrations of each row. We subtracted a linear baseline from each
calibrated spectrum and then corrected for the atmospheric opacity
using GBT weather
models.\footnote{http://www.gb.nrao.edu/~rmaddale/Weather/}




The aperture efficiencies ($\eta_a$) were derived using observations
of sources from the CARMA CALfind database. CARMA typically measured
fluxes at 103.5 GHz, but the difference between 103.5~GHz and 88~GHz
flux densities will be only $\sim 20\%$ even for a steep spectral
index ($F_{\nu} \propto \nu^{-1}$). Applying the derived aperture
efficiency to pointing source observations for each observing block
yielded consistent flux densities. The structures we observe are
comparable to the beam size, so we convert our final maps from
$T_A^\prime$ to $T_{mb}$ using the main beam efficiency ($\eta_{mb}$),
which is $\approx 1.3 \eta_a$ for the GBT \citep{gbtmemo276}. The
value of $\eta_{mb}/\eta_{a}$ changes by less than 15\% for sources
sizes up to 60\arcsec\ because of the clean nature of the GBT beam
(R.\ Maddalena, priv.~comm., 2013).


The calibrated spectra were gridded using a Bessel function tapered by
a Gaussian \citep[for details see][]{2007A&A...474..679M}.  The final
cubes have a beam size $\approx 9.2$\arcsec, slightly larger than the
native 8.3\arcsec\ GBT resolution, and were smoothed to a spectral
resolution of 3.9~MHz (13.2~\kms\ for HCN and 13.14~\kms\ for
\hcop). The typical noise per channel ($T_{mb}$) is 42mK (\hcn) and
31mK (\hcop) and each map has $\sim~100$ independent resolution
elements.

We used IRAM 30\,m \hcn\ observations of M82 from the literature to
assess the accuracy of our flux density scale.  After smoothing our
cube to the resolution of the 30\,m at 89~GHz (28\arcsec), we derive
an integrated flux of $28.75\pm1.24 \, \Kkms$ for the center of
M82. This value agrees with 30\,m \hcn\ integrated fluxes from the
literature: 29\Kkms\ \citep{2008ApJ...677..262K}, 29.9\Kkms\
\citep{2008A&A...492..675F}, 27.4\Kkms\ \citep{2004ApJS..152...63G},
and 25.4\Kkms\ \citep{1989A&A...220...57N}.



\section{Results} \label{sec:results}

\begin{figure*} \centering
\includegraphics[width=\textwidth]{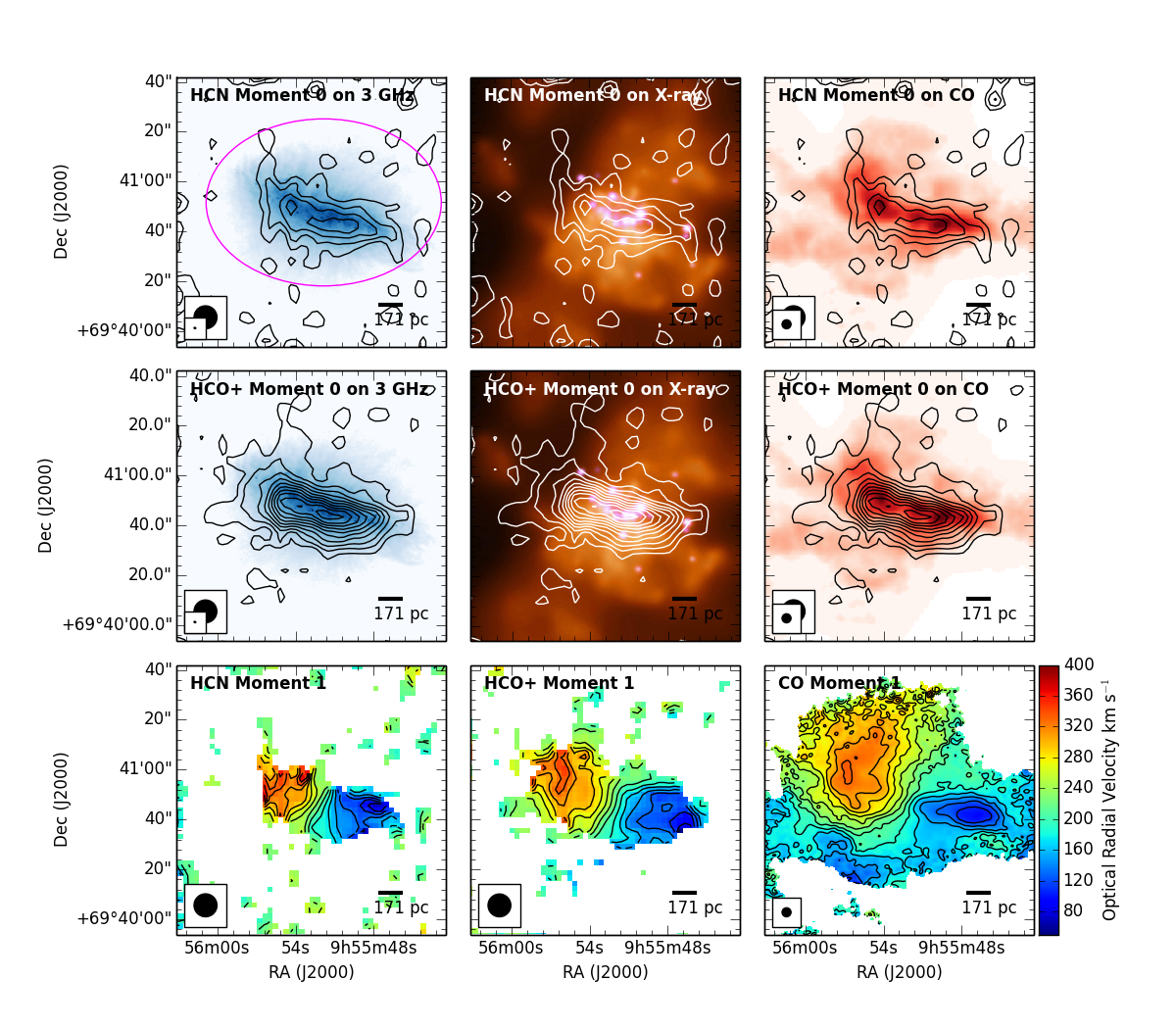}
\caption{ {\em Top and Middle Rows:} The \hcn\ (top) and \hcop\
  (middle) integrated line flux (moment zero) contours overlaid on a
  3~GHz radio continuum image (Marvil et al.\ in prep; left), an X-ray
  image with the point sources highlighted in pink
  (NASA/CXC/SAO/PSU/CMU; middle), and a \co{12}{1}{0} integrated flux
  image (\citealp{2002ApJ...580L..21W}; right).  The \hcn\ and \hcop\
  emission are both correlated with the total molecular gas and star
  formation in the center of M82. The diffuse \hcop\ emission on the
  northeastern edge of the disk correlates with the outflow seen in
  \co{12}{1}{0} by \citet{2002ApJ...580L..21W} and outlines the
  eastern edge of the hot gas associated with the outflow seen in diffuse
  X-rays \citep{2003MNRAS.343L..47S}.  The contours start at 6$\sigma$
  and go up by factors of 2 (1$\sigma_{\hcn}$=3.09 $\Kkms$ and
  1$\sigma_{\hcop}$=2.26 $\Kkms$).  The magenta ellipse in the top
  left panel indicates the region from which the spectra in
  Figure~\ref{fig:M82_all_spectra_mean} were extracted. {\em Bottom
    Row:} The first moment map (mean velocity) for \hcn, \hcop, and
  \co{12}{1}{0}. Both velocity fields are consistent with a rotating
  torus of molecular material
  \citep{1987PASJ...39..685N,1995ApJ...445L..99S}. The possibly
  outflowing \hcop\ material north and south of the disk has
  velocities similar to the outflowing material in the center of M82
  instead of velocities associated with the rotating molecular
  torus. The GBT, OVRO, and VLA beams are shown in the lower left
  corner of relevant panels and a 10\arcsec\ scale-bar in the lower
  right hand corner of all panels. }
\label{fig:M82_overview_figure}
\end{figure*}


The \hcn\ and \hcop\ maps demonstrate the GBT's excellent capabilities
for mapping large areas at high surface brightness sensitivity with
good resolution (Figure~\ref{fig:M82_overview_figure}).  Higher
resolution maps have been published for both \hcn\ $J=1-0$
\citep{1993A&A...277..381B} and \hcop\ $J=1-0$
\citep{1998ApJ...507..745S}, but our maps have better surface
brightness sensitivity by a factor of four and cover a larger area.

We compare our maps to lower resolution \co{12}{2}{1} mapping from the
HERACLES survey (Leroy et al., in prep) and to higher resolution,
zero-spacing corrected \co{12}{1}{0} interferometeric maps
\citep{2002ApJ...580L..21W}. The \co{12}{2}{1} map has 20.1\arcsec\
resolution and covers a wide field. The \co{12}{1}{0} map has higher
resolution (3.6\arcsec), but covers a smaller field of view. We also
compare our data to a new 3.0~GHz Jansky Very Large Array (VLA)
continuum map (Marvil et al., 2013, in prep), which has $\approx 0.7
\arcsec$ resolution and includes a mixture of free-free and
synchrotron emission, which both trace the distribution of recent star
formation. The total flux in the map agrees with the total flux
measured from single dish observations in the literature.

\subsection{\hcn, \hcop, and CO comparison}

The morphology of \hcn\ and \hcop\ emission follows the galaxy disk
and coincides with the main ridge of \co{12}{1}{0} emission and the
star formation traced by the radio continuum emission (top and middle
panels of Figure~\ref{fig:M82_overview_figure}).  The velocity
distributions of the disk \hcn\ and \hcop\ emission are similar to the
velocity distribution of the CO disk (bottom panels of
Figure~\ref{fig:M82_overview_figure}), suggesting that the \hcn,
\hcop, and CO emission in the disk originates from the same rotating
molecular torus \citep{1987PASJ...39..685N,1995ApJ...445L..99S}.


For the first time, we also measure \hcop\ emission at low surface
brightness associated with the \co{12}{1}{0} emission extending north
and south of the main disk \citep{2002ApJ...580L..21W}. Simulations of
the GBT beam shape at 4\,mm indicate that this emission does not
originate from sidelobes. The \hcop\ emission outlines the eastern
edge of the X-ray emission associated with the central outflow,
suggesting that the dense molecular gas is entrained in the outflow of
lower density gas. The \hcop\ in this region also is kinematically
inconsistent with a disk, similar to what is seen in CO (cf.\ Figure 6
in \citealp{2002ApJ...580L..21W} and
Figure~\ref{fig:M82_overview_figure} here).



CO emission has been associated with the outflow in M82
\citep{2002ApJ...580L..21W} and has been seen in the outflow from the
starburst nucleus of NGC 253 \citep{2013Natur.499..450B}. Emission
from \hcn\ and \hcop\ has also been associated with the AGN-driven
outflow in the ULIRG Mrk 231 \citep{2012A&A...537A..44A}. To our
knowledge, the current observations, however, would be the first time
that dense molecular gas, as traced by \hcop, has been found to be
associated with a starburst-driven outflow in a nearby galaxy. The
outflow of dense molecular gas seen in \hcop\ may regulate star
formation in galaxies like M82 by removing the fuel for star
formation.



\begin{figure} 
\centering
\includegraphics[width=\columnwidth]{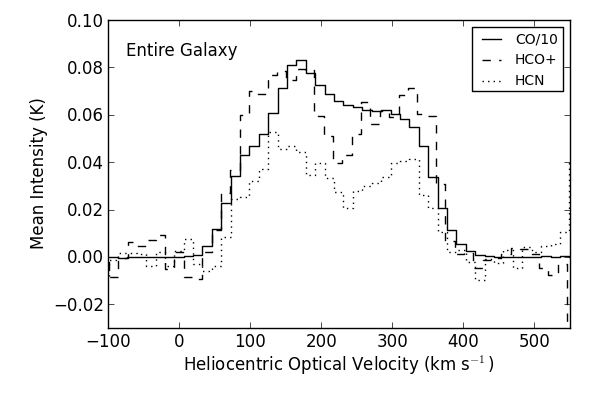}
\caption{Mean intensity averaged over the disk of M82 
  for \co{12}{2}{1} (Leroy et al., in prep), \hcn, and \hcop. The
  region averaged is shown as an magenta
  ellipse in top left panel of
  Figure~\ref{fig:M82_overview_figure}. The CO intensity has been
  divided by 10.
}
\label{fig:M82_all_spectra_mean}
\end{figure}

The disk-averaged line profiles of the \hcn\ and \hcop\ emission agree
with \co{12}{2}{1} emission from the HERACLES image (Leroy et al., in
prep) in Figure~\ref{fig:M82_all_spectra_mean}.  Both \hcn\ and \hcop\
have structure near 220~\kms, which is not seen in \co{12}{2}{1} but
is seen in the higher order CO transitions
\citep{2010A&A...521L...2L}.  The models of the $\rm{CO}$ emission
from \citet{2010A&A...521L...2L} show that the different CO
transitions reflect different molecular gas densities. Transitions
like $\co{12}{2}{1}$ are emitted by relatively diffuse ($10^{3.5} \,
\cc$) molecular gas associated with the disk, while the higher order
$\rm{CO}$ transitions ($J > 4$) are emitted by two denser components
($10^5\, \cc$ and $10^6\, \cc$) associated with the star-forming gas
\citep{2010A&A...521L...2L}.  Inspection of the channel maps near
220~\kms\ confirm that the lack of structure in the \co{12}{2}{1} is
due to the presence of a significant amount of CO emission associated
with the warm and diffuse molecular gas found throughout the disk,
while the \hcn\ and \hcop\ trace the denser molecular gas component
found in the torus.




\subsection{The Relationship Between \hcn, \hcop, CO, and Star
  Formation}

To explore the relationships between \hcn, \hcop, CO, and star
formation within M82, we smoothed the images to the same resolution
(9.2\arcsec), regridded them to the same coordinate system, and
rebinned the pixels so that each pixel represents an independent
sample. The line intensities were derived from
moment zero maps and the luminosities were calculated by multiplying
the intensities by the area of a pixel. Regions less than 3$\sigma$
were blanked.

\begin{figure*}
\centering
\includegraphics[width=\textwidth]{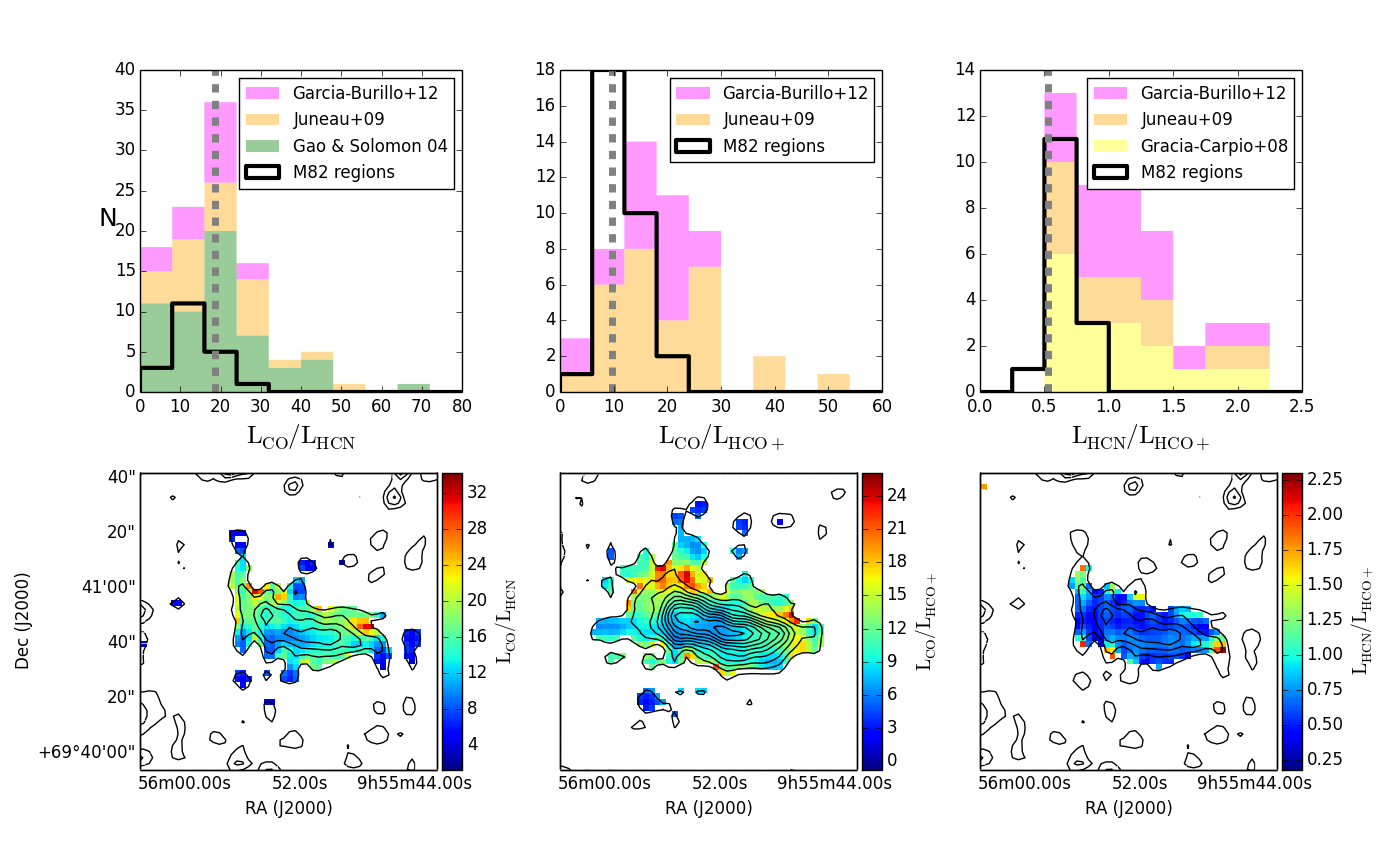}
\caption{{\em Top:} The distributions of \LCOHCN\ (left), \LCOHCOP\
  (middle), and \LHCNHCOP\ (right) for regions in our M82 data
  compared to values for entire galaxies or galaxy centers from the
  literature
  \citep{2004ApJ...606..271G,2008A&A...479..703G,2009ApJ...707.1217J,2012A&A...539A...8G}. For
  M82, only ratios with S/N greater than 6 are shown. The ratios
  between the total values for each quantity for M82 are shown as
  dashed gray lines.  The ratios for M82 are similar to those seen in
  the literature, although the M82 distributions peak at lower ratios
  than the bulk of the literature values. {\em Bottom:} Spatial
  distribution of the \LCOHCN\ (left), \LCOHCOP\ (middle), and
  \LHCNHCOP\ (right) ratios with the \hcn\ (left and right) and \hcop\
  (middle) contours from Figure~\ref{fig:M82_overview_figure}. The
  \LCOHCN\ and \LCOHCOP\ values increase away from the center of the
  galaxy reflecting a decrease in dense gas away from the center of
  the starburst. The \LHCNHCOP\ ratio is constant across the face of
  the galaxy with slightly higher values found in the southern half of
  the galaxy. }
\label{fig:L_HCN_HCOP_ratio}
\end{figure*}

Figure~\ref{fig:L_HCN_HCOP_ratio} compares the distribution of
\LCOHCN, \LCOHCOP, and \LHCNHCOP\ for regions within M82 and for
entire galaxies (or centers of galaxies) from the literature
\citep{2004ApJ...606..271G,2008A&A...479..703G,2009ApJ...707.1217J,2012A&A...539A...8G}.
In M82, the distributions of all three ratios have the same range as
the points from the literature, although \LCOHCN\ and \LCOHCOP\ do
peak at lower values. However, we must be cautious because we are
comparing measurements of entire galaxies with spatially resolved
measurements within a galaxy. Because the CO emission has a larger
filling factor than the HCN emission, the unresolved measurements may
have systematically larger \LCOHCN\ and \LCOHCOP\ ratios. The \LCOHCN\
and \LHCNHCOP\ values for M82 as a whole are similar to the mode of
the values found for other galaxies, while the \LCOHCOP\ value is
slightly lower. The \LCOHCN\ and \LCOHCOP\ values increase with
distance from the center of the galaxy (bottom panels of
Figure~\ref{fig:L_HCN_HCOP_ratio}), suggesting the fraction of dense
gas decreases with distance from the center. The \LHCNHCOP\ ratio is
roughly constant across the center of the galaxy with higher values at
the southern edge.




\begin{figure*}
\centering
\includegraphics[width=\textwidth]{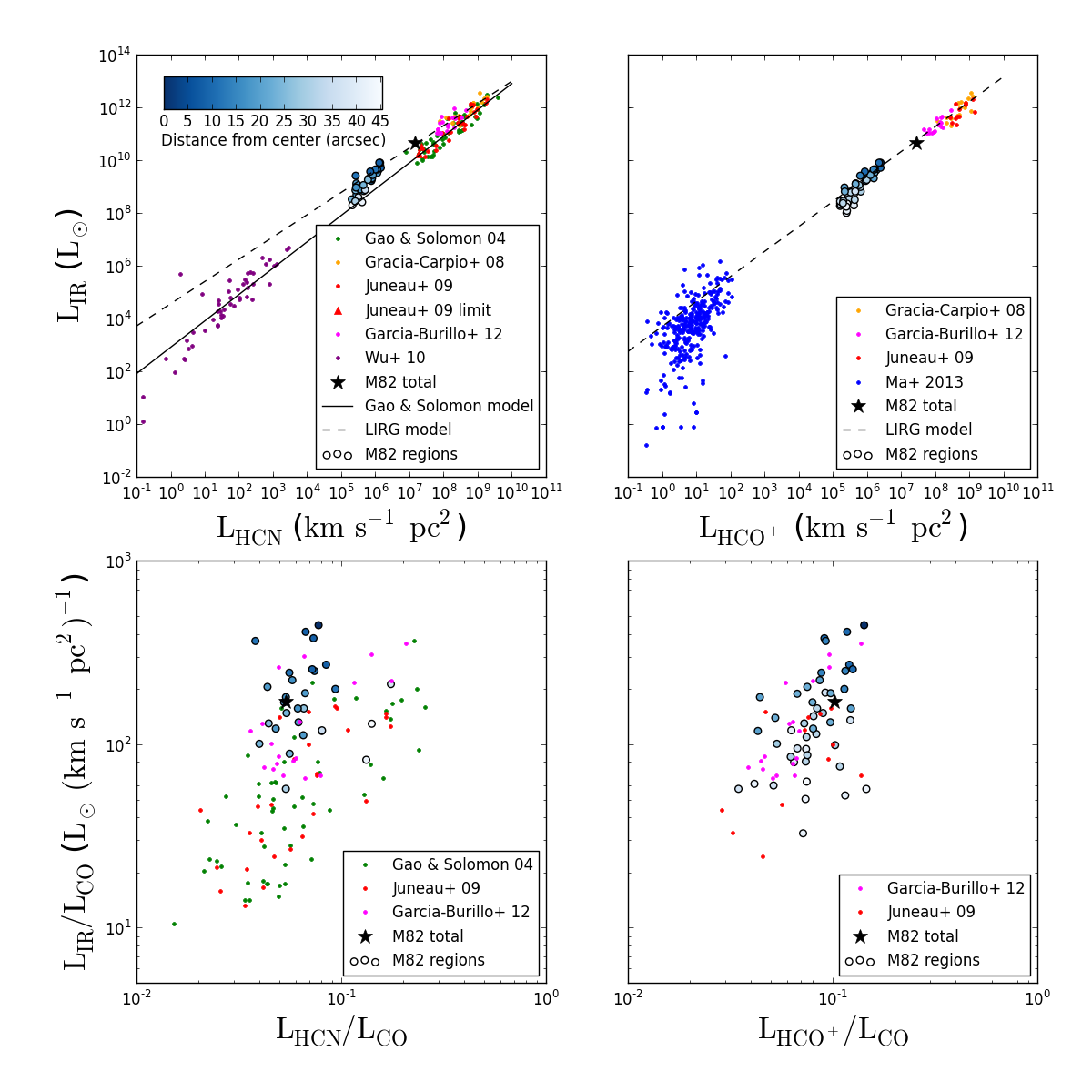}
\caption{ {\em Top:} The star formation rate traced by infrared
  luminosity as a function of the amount of dense gas traced by \hcn\
  (left) and \hcop\ (right). The \LIR-\LHCN\ fit for a sample
  including both normal galaxies and LIRGs/ULIRGs from
  \citet{2004ApJS..152...63G} is shown as a solid line; star-forming
  regions within the Milky Way also follow this fit
  \citep{2010ApJS..188..313W,2012arXiv1211.6492M}. The dotted line
  shows the \LIR-\LHCN\ fit derived from the sample of LIRGs/ULIRGs
  \citep{2008A&A...479..703G,2012A&A...539A...8G}.  The integrated
  \LIR\ and \LHCN\ points for M82 follow the LIRG/ULIRG relationship
  between \LIR\ and \LHCN.  Regions within M82, however, span a range
  of values with the high luminosity \hcn\ and \hcop\ points following
  the LIRG/ULIRG relationship and the low luminosity points following
  instead the relationship seen in normal galaxies and in the Milky
  Way. We see a similar trend for \LIR\ vs. \LHCOP.  {\em Bottom:} In
  M82, the amount of dense gas per total molecular gas mass (\LHCNCO\
  and \LHCOPCO) and the amount of star formation per total molecular
  gas mass (\LIR/\LCO) are both high.  The errors on the M82 data in
  all plots are smaller than the symbol size.}
\label{fig:gao_and_solomon}
\end{figure*}

Figure~\ref{fig:gao_and_solomon} compares the relationship between the
total molecular gas mass (\LCO), the dense gas mass (\LHCN\ and
\LHCOP), and the star formation rate (\LIR) for the entire M82 disk,
points within M82, the sample of galaxies from the literature used in
Figure~\ref{fig:L_HCN_HCOP_ratio}, and star-forming regions in the
Milky Way \citep{2010ApJS..188..313W,2012arXiv1211.6492M}. We have
estimated \LIR\ for the M82 points by multiplying the 3.0~GHz
continuum flux density per pixel by the ratio of the infrared
luminosity to the 3~GHz flux density for the entire galaxy. In effect,
this procedure relies on the radio-infrared correlation, one of the
tightest astronomical correlations, but avoids additional systematic
errors by using the empirical ratio rather than a fit to the
correlation seen in a large sample of galaxies. The use of the 3~GHz
radio continuum ameliorates the significant optical depth effects
found in edge-on galaxies like M82.

For the entire M82 disk, the relationship between the \LIR\ and \LHCN\
values matches the trend between \LIR\ and \LHCN\ found for a sample
of LIRGs and ULIRGs, which have high star formation rates. However,
our data shows that the relationship between \LIR\ and \LHCN\ varies
{\em within M82}. Regions away from the central starburst tend to have
lower \LIR\ (star formation rate) for a given amount of \LHCN\ (dense
molecular gas). These points match the trend seen in normal galaxies
\citep{2004ApJ...606..271G} and individual star-forming regions in the
Milky Way \citep{2010ApJS..188..313W}.  For points near the central
starburst, the \LIR\ (star formation rate) is higher for a given
amount of \LHCN\ (dense molecular gas), matching the trend seen for
LIRG/ULIRGs. We see a similar trend for the \hcop\
measurements. Compared to normal galaxies, individual regions in M82
tend to have higher dense gas fractions (\LHCNCO\ or \LHCOPCO) and
higher ratios of star formation to total molecular gas mass
($\LIR/\LCO$), but the dense gas fractions vary by a factor of 10.

These resolved observations of M82 show that the relationship between
the amount of dense gas and star formation rate (as traced by the
radio continuum) varies within a single galaxy. The key variable
appears to be the central starburst, which could be using up or
expelling the dense gas, affecting the gas tracer chemistry, and/or
changing how star formation proceeds. Future resolved observations of
\hcn\ and \hcop\ in star-forming regions in a variety of galactic
environments will allow us to disentangle these possibilities and
understand how the central starburst in M82 influences star formation.

\section{Summary} \label{sec:conclusions}

We have made the most sensitive map to date of the dense molecular gas
in the starburst galaxy M82 using the largest single-dish
millimeter-wave telescope in the world: the GBT. The \hcn\ and \hcop\
emission correlates with lower density molecular gas, traced by CO,
and star formation, traced by radio continuum, and its kinematics are
consistent with the previously proposed torus of molecular gas. We
also detect low surface brightness \hcop\ emission coincident with the
base of the molecular gas outflow first detected in $\rm{CO}$ and
tracing the edge of the hot outflowing gas seen in the X-ray.

The \LCOHCN, \LCOHCOP, and \LHCNHCOP\ ratios are similar to those in
other starburst galaxies. The first two ratios increase with distance
from the central starburst, implying that the fraction of dense gas
decreases with distance from the starburst.

The relationship between the dense molecular gas and star formation
varies with distance from the central starburst. Near its center,
there is a higher ratio of star formation to dense molecular gas,
similar to the relationship seen for LIRGs and ULIRGs, but outside of
the central starburst, the ratio of the star formation to dense
molecular gas decreases, agreeing with the correlation seen in normal
galaxies and the Milky Way.

These observations demonstrate the effectiveness of the GBT for
mapping dense molecular gas in external galaxies. The already-exciting
capabilities of the GBT will be increased further with the advent of
the 16 element 4mm feed array (ARGUS) being built for the GBT and will
complement on-going efforts with ALMA.

\acknowledgments The National Radio Astronomy Observatory is a
facility of the National Science Foundation operated under cooperative
agreement by Associated Universities, Inc.  This research used APLpy
and Astropy \citep{2013arXiv1307.6212T}.

{\it Facilities:} \facility{GBT, VLA}


\begin{thebibliography}{30}
\expandafter\ifx\csname natexlab\endcsname\relax\def\natexlab#1{#1}\fi

\bibitem[{{Aalto} {et~al.}(2012){Aalto}, {Garcia-Burillo}, {Muller}, {Winters},
  {van der Werf}, {Henkel}, {Costagliola}, \& {Neri}}]{2012A&A...537A..44A}
{Aalto}, S., {Garcia-Burillo}, S., {Muller}, S., {Winters}, J.~M., {van der
  Werf}, P., {Henkel}, C., {Costagliola}, F., \& {Neri}, R. 2012, \aap, 537,
  A44

\bibitem[{{Bolatto} {et~al.}(2013){Bolatto}, {Warren}, {Leroy}, {Walter},
  {Veilleux}, {Ostriker}, {Ott}, {Zwaan}, {Fisher}, {Weiss}, {Rosolowsky}, \&
  {Hodge}}]{2013Natur.499..450B}
{Bolatto}, A.~D., {Warren}, S.~R., {Leroy}, A.~K., {Walter}, F., {Veilleux},
  S., {Ostriker}, E.~C., {Ott}, J., {Zwaan}, M., {Fisher}, D.~B., {Weiss}, A.,
  {Rosolowsky}, E., \& {Hodge}, J. 2013, \nat, 499, 450

\bibitem[{{Brouillet} \& {Schilke}(1993)}]{1993A&A...277..381B}
{Brouillet}, N., \& {Schilke}, P. 1993, \aap, 277, 381

\bibitem[{{Carilli} \& {Walter}(2013)}]{2013ARA&A..51..105C}
{Carilli}, C.~L., \& {Walter}, F. 2013, \araa, 51, 105

\bibitem[{{Dalcanton} {et~al.}(2009){Dalcanton}, {Williams}, {Seth}, {Dolphin},
  {Holtzman}, {Rosema}, {Skillman}, {Cole}, {Girardi}, {Gogarten},
  {Karachentsev}, {Olsen}, {Weisz}, {Christensen}, {Freeman}, {Gilbert},
  {Gallart}, {Harris}, {Hodge}, {de Jong}, {Karachentseva}, {Mateo}, {Stetson},
  {Tavarez}, {Zaritsky}, {Governato}, \& {Quinn}}]{2009ApJS..183...67D}
{Dalcanton}, J.~J., {Williams}, B.~F., {Seth}, A.~C., {Dolphin}, A.,
  {Holtzman}, J., {Rosema}, K., {Skillman}, E.~D., {Cole}, A., {Girardi}, L.,
  {Gogarten}, S.~M., {Karachentsev}, I.~D., {Olsen}, K., {Weisz}, D.,
  {Christensen}, C., {Freeman}, K., {Gilbert}, K., {Gallart}, C., {Harris}, J.,
  {Hodge}, P., {de Jong}, R.~S., {Karachentseva}, V., {Mateo}, M., {Stetson},
  P.~B., {Tavarez}, M., {Zaritsky}, D., {Governato}, F., \& {Quinn}, T. 2009,
  \apjs, 183, 67

\bibitem[{{Fuente} {et~al.}(2008){Fuente}, {Garc{\'{\i}}a-Burillo}, {Usero},
  {Gerin}, {Neri}, {Faure}, {Le Bourlot}, {Gonz{\'a}lez-Garc{\'{\i}}a},
  {Rizzo}, {Alonso-Albi}, \& {Tennyson}}]{2008A&A...492..675F}
{Fuente}, A., {Garc{\'{\i}}a-Burillo}, S., {Usero}, A., {Gerin}, M., {Neri},
  R., {Faure}, A., {Le Bourlot}, J., {Gonz{\'a}lez-Garc{\'{\i}}a}, M., {Rizzo},
  J.~R., {Alonso-Albi}, T., \& {Tennyson}, J. 2008, \aap, 492, 675

\bibitem[{{Gao} \& {Solomon}(2004{\natexlab{a}})}]{2004ApJS..152...63G}
{Gao}, Y., \& {Solomon}, P.~M. 2004{\natexlab{a}}, \apjs, 152, 63

\bibitem[{{Gao} \& {Solomon}(2004{\natexlab{b}})}]{2004ApJ...606..271G}
---. 2004{\natexlab{b}}, \apj, 606, 271

\bibitem[{{Garc{\'{\i}}a-Burillo} {et~al.}(2012){Garc{\'{\i}}a-Burillo},
  {Usero}, {Alonso-Herrero}, {Graci{\'a}-Carpio}, {Pereira-Santaella},
  {Colina}, {Planesas}, \& {Arribas}}]{2012A&A...539A...8G}
{Garc{\'{\i}}a-Burillo}, S., {Usero}, A., {Alonso-Herrero}, A.,
  {Graci{\'a}-Carpio}, J., {Pereira-Santaella}, M., {Colina}, L., {Planesas},
  P., \& {Arribas}, S. 2012, \aap, 539, A8

\bibitem[{{Graci{\'a}-Carpio} {et~al.}(2008){Graci{\'a}-Carpio},
  {Garc{\'{\i}}a-Burillo}, {Planesas}, {Fuente}, \&
  {Usero}}]{2008A&A...479..703G}
{Graci{\'a}-Carpio}, J., {Garc{\'{\i}}a-Burillo}, S., {Planesas}, P., {Fuente},
  A., \& {Usero}, A. 2008, \aap, 479, 703

\bibitem[{{Heiderman} {et~al.}(2010){Heiderman}, {Evans}, {Allen}, {Huard}, \&
  {Heyer}}]{2010ApJ...723.1019H}
{Heiderman}, A., {Evans}, II, N.~J., {Allen}, L.~E., {Huard}, T., \& {Heyer},
  M. 2010, \apj, 723, 1019

\bibitem[{{Juneau} {et~al.}(2009){Juneau}, {Narayanan}, {Moustakas}, {Shirley},
  {Bussmann}, {Kennicutt}, \& {Vanden Bout}}]{2009ApJ...707.1217J}
{Juneau}, S., {Narayanan}, D.~T., {Moustakas}, J., {Shirley}, Y.~L.,
  {Bussmann}, R.~S., {Kennicutt}, Jr., R.~C., \& {Vanden Bout}, P.~A. 2009,
  \apj, 707, 1217

\bibitem[{{Kainulainen} {et~al.}(2009){Kainulainen}, {Beuther}, {Henning}, \&
  {Plume}}]{2009A&A...508L..35K}
{Kainulainen}, J., {Beuther}, H., {Henning}, T., \& {Plume}, R. 2009, \aap,
  508, L35

\bibitem[{{Krips} {et~al.}(2008){Krips}, {Neri}, {Garc{\'{\i}}a-Burillo},
  {Mart{\'{\i}}n}, {Combes}, {Graci{\'a}-Carpio}, \&
  {Eckart}}]{2008ApJ...677..262K}
{Krips}, M., {Neri}, R., {Garc{\'{\i}}a-Burillo}, S., {Mart{\'{\i}}n}, S.,
  {Combes}, F., {Graci{\'a}-Carpio}, J., \& {Eckart}, A. 2008, \apj, 677, 262

\bibitem[{{Lada} {et~al.}(2012){Lada}, {Forbrich}, {Lombardi}, \&
  {Alves}}]{2012ApJ...745..190L}
{Lada}, C.~J., {Forbrich}, J., {Lombardi}, M., \& {Alves}, J.~F. 2012, \apj,
  745, 190

\bibitem[{{Lada} {et~al.}(2010){Lada}, {Lombardi}, \&
  {Alves}}]{2010ApJ...724..687L}
{Lada}, C.~J., {Lombardi}, M., \& {Alves}, J.~F. 2010, \apj, 724, 687

\bibitem[{{Leroy} {et~al.}(2013){Leroy}, {Walter}, {Sandstrom}, {Schruba},
  {Munoz-Mateos}, {Bigiel}, {Bolatto}, {Brinks}, {de Blok}, {Meidt}, {Rix},
  {Rosolowsky}, {Schinnerer}, {Schuster}, \& {Usero}}]{2013AJ....146...19L}
{Leroy}, A.~K., {Walter}, F., {Sandstrom}, K., {Schruba}, A., {Munoz-Mateos},
  J.-C., {Bigiel}, F., {Bolatto}, A., {Brinks}, E., {de Blok}, W.~J.~G.,
  {Meidt}, S., {Rix}, H.-W., {Rosolowsky}, E., {Schinnerer}, E., {Schuster},
  K.-F., \& {Usero}, A. 2013, \aj, 146, 19

\bibitem[{{Loenen} {et~al.}(2010){Loenen}, {van der Werf}, {G{\"u}sten},
  {Meijerink}, {Israel}, {Requena-Torres}, {Garc{\'{\i}}a-Burillo}, {Harris},
  {Klein}, {Kramer}, {Lord}, {Mart{\'{\i}}n-Pintado}, {R{\"o}llig}, {Stutzki},
  {Szczerba}, {Wei{\ss}}, {Philipp-May}, {Yorke}, {Caux}, {Delforge},
  {Helmich}, {Lorenzani}, {Morris}, {Philips}, {Risacher}, \&
  {Tielens}}]{2010A&A...521L...2L}
{Loenen}, A.~F., {van der Werf}, P.~P., {G{\"u}sten}, R., {Meijerink}, R.,
  {Israel}, F.~P., {Requena-Torres}, M.~A., {Garc{\'{\i}}a-Burillo}, S.,
  {Harris}, A.~I., {Klein}, T., {Kramer}, C., {Lord}, S.,
  {Mart{\'{\i}}n-Pintado}, J., {R{\"o}llig}, M., {Stutzki}, J., {Szczerba}, R.,
  {Wei{\ss}}, A., {Philipp-May}, S., {Yorke}, H., {Caux}, E., {Delforge}, B.,
  {Helmich}, F., {Lorenzani}, A., {Morris}, P., {Philips}, T.~G., {Risacher},
  C., \& {Tielens}, A.~G.~G.~M. 2010, \aap, 521, L2

\bibitem[{{Ma} {et~al.}(2012){Ma}, {Tan}, \& {Barnes}}]{2012arXiv1211.6492M}
{Ma}, B., {Tan}, J.~C., \& {Barnes}, P.~J. 2012, ArXiv e-prints

\bibitem[{{Maddalena}(2010)}]{gbtmemo276}
{Maddalena}, R.~J. 2010, {Theoretical Ratio of Beam Efficiency to Aperture
  Efficiency}, GBT Memo 276, {National Radio Astronomy Observatory}

\bibitem[{{Mangum} {et~al.}(2007){Mangum}, {Emerson}, \&
  {Greisen}}]{2007A&A...474..679M}
{Mangum}, J.~G., {Emerson}, D.~T., \& {Greisen}, E.~W. 2007, \aap, 474, 679

\bibitem[{{Nakai} {et~al.}(1987){Nakai}, {Hayashi}, {Handa}, {Sofue},
  {Hasegawa}, \& {Sasaki}}]{1987PASJ...39..685N}
{Nakai}, N., {Hayashi}, M., {Handa}, T., {Sofue}, Y., {Hasegawa}, T., \&
  {Sasaki}, M. 1987, \pasj, 39, 685

\bibitem[{{Nguyen-Q-Rieu} {et~al.}(1989){Nguyen-Q-Rieu}, {Nakai}, \&
  {Jackson}}]{1989A&A...220...57N}
{Nguyen-Q-Rieu}, {Nakai}, N., \& {Jackson}, J.~M. 1989, \aap, 220, 57

\bibitem[{{Rieke} {et~al.}(1980){Rieke}, {Lebofsky}, {Thompson}, {Low}, \&
  {Tokunaga}}]{1980ApJ...238...24R}
{Rieke}, G.~H., {Lebofsky}, M.~J., {Thompson}, R.~I., {Low}, F.~J., \&
  {Tokunaga}, A.~T. 1980, \apj, 238, 24

\bibitem[{{Seaquist} {et~al.}(1998){Seaquist}, {Frayer}, \&
  {Bell}}]{1998ApJ...507..745S}
{Seaquist}, E.~R., {Frayer}, D.~T., \& {Bell}, M.~B. 1998, \apj, 507, 745

\bibitem[{{Shen} \& {Lo}(1995)}]{1995ApJ...445L..99S}
{Shen}, J., \& {Lo}, K.~Y. 1995, \apjl, 445, L99

\bibitem[{{Stevens} {et~al.}(2003){Stevens}, {Read}, \&
  {Bravo-Guerrero}}]{2003MNRAS.343L..47S}
{Stevens}, I.~R., {Read}, A.~M., \& {Bravo-Guerrero}, J. 2003, \mnras, 343, L47

\bibitem[{{The Astropy Collaboration} {et~al.}(2013){The Astropy
  Collaboration}, {Robitaille}, {Tollerud}, {Greenfield}, {Droettboom}, {Bray},
  {Aldcroft}, {Davis}, {Ginsburg}, {Price-Whelan}, {Kerzendorf}, {Conley},
  {Crighton}, {Barbary}, {Muna}, {Ferguson}, {Grollier}, {Parikh}, {Nair},
  {G{\"u}nther}, {Deil}, {Woillez}, {Conseil}, {Kramer}, {Turner}, {Singer},
  {Fox}, {Weaver}, {Zabalza}, {Edwards}, {Azalee Bostroem}, {Burke}, {Casey},
  {Crawford}, {Dencheva}, {Ely}, {Jenness}, {Labrie}, {Lian Lim},
  {Pierfederici}, {Pontzen}, {Ptak}, {Refsdal}, {Servillat}, \&
  {Streicher}}]{2013arXiv1307.6212T}
{The Astropy Collaboration}, {Robitaille}, T.~P., {Tollerud}, E.~J.,
  {Greenfield}, P., {Droettboom}, M., {Bray}, E., {Aldcroft}, T., {Davis}, M.,
  {Ginsburg}, A., {Price-Whelan}, A.~M., {Kerzendorf}, W.~E., {Conley}, A.,
  {Crighton}, N., {Barbary}, K., {Muna}, D., {Ferguson}, H., {Grollier}, F.,
  {Parikh}, M.~M., {Nair}, P.~H., {G{\"u}nther}, H.~M., {Deil}, C., {Woillez},
  J., {Conseil}, S., {Kramer}, R., {Turner}, J.~E.~H., {Singer}, L., {Fox}, R.,
  {Weaver}, B.~A., {Zabalza}, V., {Edwards}, Z.~I., {Azalee Bostroem}, K.,
  {Burke}, D.~J., {Casey}, A.~R., {Crawford}, S.~M., {Dencheva}, N., {Ely}, J.,
  {Jenness}, T., {Labrie}, K., {Lian Lim}, P., {Pierfederici}, F., {Pontzen},
  A., {Ptak}, A., {Refsdal}, B., {Servillat}, M., \& {Streicher}, O. 2013,
  ArXiv e-prints

\bibitem[{{Walter} {et~al.}(2002){Walter}, {Weiss}, \&
  {Scoville}}]{2002ApJ...580L..21W}
{Walter}, F., {Weiss}, A., \& {Scoville}, N. 2002, \apjl, 580, L21

\bibitem[{{Wu} {et~al.}(2010){Wu}, {Evans}, {Shirley}, \&
  {Knez}}]{2010ApJS..188..313W}
{Wu}, J., {Evans}, II, N.~J., {Shirley}, Y.~L., \& {Knez}, C. 2010, \apjs, 188,
  313

\end{thebibliography}

\end{document}